     [1999/12/01 v1.4c Il Nuovo Cimento]
\documentclass{cimento}
\usepackage{graphicx}
\usepackage{amsmath}
\title{A \textit{gedankenexperiment} in gravitation}
\author{Y.~Gaspar\from{ins:x}\ETC,
G.~Acquaviva\from{ins:y},
       }
\instlist{\inst{ins:x} Dipartimento di Matematica e Fisica, Universit\`{a} Cattolica del Sacro Cuore, Via Trieste 17, 25121 Brescia (BS),  
Italy;\  y.gaspar@dmf.unicatt.it
  \inst{ins:y} Dipartimento di Matematica e Fisica, Universit\`{a} degli Studi di Trento, via Sommarive 14, 38123 Povo (TN), Italy;\ acquaviva@science.unitn.it}
\PACSes{\PACSit{04.90.+e}{} \PACSit{04.20.Cv}{}}
\begin{document}

\maketitle

\begin{abstract}
In this paper we consider a thought experiment involving the effect of gravitation on an ideal scale containing a  photon. If the tidal forces inherent to a  gravitational field are neglected, then one is led to a scenario which seems to bring about perpetual motion violating the first and the second principle of thermodynamics. The tidal effects of gravitation must necessarily be included in order to obtain a consistent physical theory. As a result, Albert Einstein's thought experiments according to which the physical effects of inertial forces in an accelerated reference frame are equivalent to the effects of gravity in a frame at rest on the surface of a massive body must be reconsidered, since linearly accelerated frames do not produce tidal effects. We argue that the equivalence between inertial effects and gravitation can be restored for rotating frames and in this context a relation with the possible quantum nature of gravity is conjectured. Furthermore, these arguments show that rotation is not  merely a kinematical fact, but an essential physical reality.
\end{abstract}

\section*{Introduction}
In order to extend or to generalise the principle of relativity from linear uniform motion to accelerated and rotational motion, the thought experiments elaborated by A. Einstein have a crucial role and on their basis the general theory of relativity has been formulated. Einstein further imagined an ideal experiment (the Einstein light-box) in order to probe the consistency of quantum mechanics \cite{dy}: in this experiment a box full of light is placed in a gravitational field. Niels Bohr ultimately showed that there is no violation of  the principles of quantum mechanics. Thought experiments continue to play an important guiding role in theoretical physics: the ideal experiment conceived by J. Bekenstein \cite{be} in order to elucidate black hole entropy or the thought experiment imagined by S. Hawking to tackle the problem of unitarity violation in black hole space-times are some examples. F. Dyson \cite{dy} argues that if no conceivable thought experiment can show effects of quantum gravity, then the latter looses its physical meaning. Thus the search for ideal experiments analysing the features of new fundamental physics theories is of great relevance.

The thought experiments we consider in this paper are formulated in the context of classical general relativity. An interesting link between thermodynamics and gravitation is shown and as a consequence a relation might exist with the Weyl Curvature Hypothesis of R. Penrose \cite{pe}. All the arguments presented in this work have a heuristic nature, typical of thought experiment analysis, and will need further rigorous developments. Throughout the  paper, the following physical principles are assumed to be true:

\begin{enumerate}
 \item The positivity of mass and the equivalence between mass and energy, as predicted by special relativity, $E=mc^2$
 \item the existence of single photons having energy proportional to frequency, $E=h \nu$ and carrying momentum $p=E/c$
 \item the equivalence between inertial and gravitational mass
 \item Newton's theory of gravitation for weak fields, Einstein's general theory of relativity and in particular the phenomenon of gravitational redshift
 \item The principles of thermodynamics.
\end{enumerate}
\

In particular, as a consequence of these principles, we assume that light rays can be bended in a gravitational field and that photons have a weight. 
In the next section we will explain the construction of the thought experiment and show how it violates the principles of thermodynamics if tidal effects of gravity are not included and that this violation would occur if one performs the experiment in a linearly accelerated frame.
In section 3 the experiment is analysed in rotating frames and in this context one is led to consider rotation not as a merely kinematical motion, but as an essential physical reality. This aspect of the present analysis agrees with the work of Gr\o{}n \cite{gr}, in which rotation is considered to be  physically distinct from other accelerated motions.   

In the conclusion some relations with aspects of quantum gravity and with Penrose's Weyl Curvature Hypothesis are conjectured and a general discussion is presented.

\section{The thought experiment}

\subsection{The structure}
Consider the system represented in fig.(1), namely a tall tower supporting a wide and frictionless scale or balance. 
\

\begin{figure}
 \begin{minipage}{190pt}
   \centering
   \includegraphics[width=160pt]{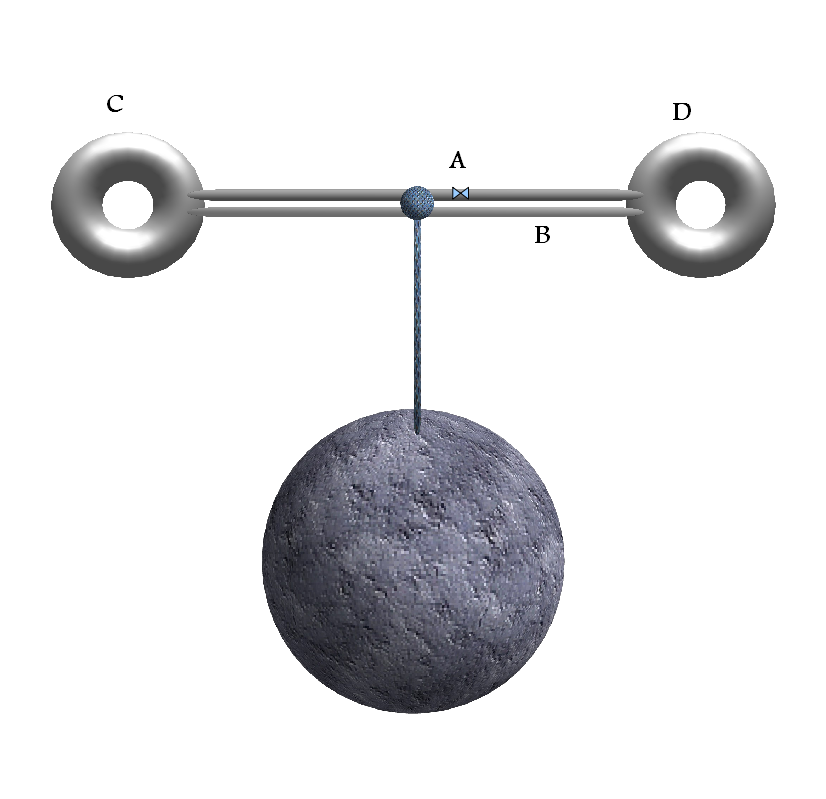}
   \caption{structure of the balance. (A) a small aperture, which allows photons to enter the (B) straight segments of the optical fiber, towards (C and D) the external loops containing coiling of optical fiber.}
 \end{minipage}
 \ \hspace{2mm} \hspace{3mm} \
 \begin{minipage}{190pt}
  \centering
   \includegraphics[width=160pt]{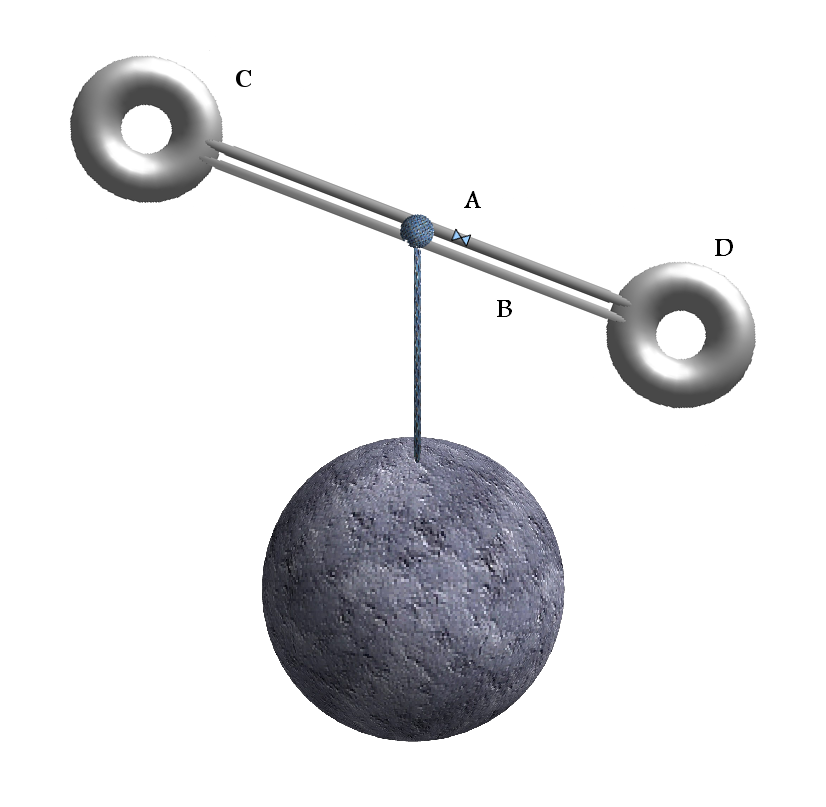}
 \caption{After a time $t_1 + t_2$, the coil D is heavier than coil C due to the presence of the photon traveling inside D.}
 \end{minipage}
\end{figure}
\

The tilting part of the balance has the following properties:
\

\begin{enumerate}
\item it is made of a perfectly reflecting optical fiber structured as a \textit{closed path}. Perfect reflection is also considered by H. Bondi in his thought experiments \cite{bo}, which shows how to avoid a violation of the first principle of thermodynamics when considering photons interacting with atoms in a gravitational field
\item a small aperture (A) near the fulcrum of the balance allows incoming photons to enter the fiber and begin to travel through it;
\item the structure of the balance is such that the time spent by these photons in the terminals (C and D) is greater than the time spent in the straight central part (B); this can be achieved, for example, by coiling  a long portion of the fiber inside the external loops.  Another way to achieve this could be to build the terminals as perfectly reflecting cavities, so that the photon will enter and will remain inside for a (statistically) long time before it finds the way out again;
\item the masses of the terminal parts are equal and very large compared to the mass of the straight central part.
\end{enumerate}
\

Now let a sufficiently localized wave packet (photon) enter the fiber through the aperture A and start to travel in it.  The structure allows the wave packet to travel back and forth inside the fiber, from a coil to the other.  Note that the ideal optical properties of the fiber prevent any dispersion that might occur to the wave packet during its journey.
\
\\

Let the photon travel along B in a time $t_{1}$ and inside the coil D in a time $t_{2} \gg t_{1}$.  Because of the equivalence between mass and energy and because of the equivalence between inertial and gravitational mass, the presence of a photon in coil D increases the gravitational mass of  that portion of the balance: as a result, the balance will undergo a tilt towards D and a consequent torque around the fulcrum.  After the time $t_2$, the situation is as represented in fig. 2.
\

Waiting a period of time $2 t_1$ after the run inside coil D, we would find the photon entering the coil C and spend another $t_2$ travelling in it, such that the gravitational mass of the C-portion will be increased and as a result, the system will undergo a movement towards the side of the C coil (see fig.3). Once out again, the photon can start a new cycle.
\
\\

\begin{center}
\begin{figure}
 \includegraphics[width=180pt]{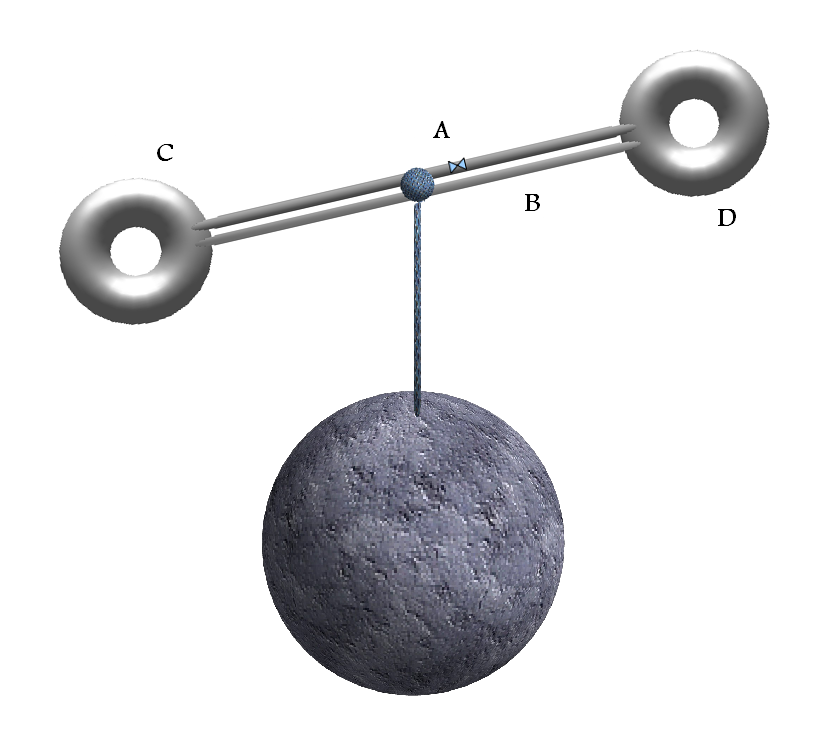}
 \caption{After a time $3t_1 + 2t_2$, the coil C is heavier than coil D due to the presence of the photon traveling inside C.}
\end{figure}
\end{center}

\subsection{Thermodynamics}
Let us now examine what happens if the experiment is performed in a linearly accelerated frame.  The equivalent gravitational field perceived by an observer in this frame has no tidal effects.  The scale or balance seems to be able to describe an infinity of oscillations around the fulcrum: indeed the cycle described so far in the previous section can repeat itself \textit{ad infinitum}.
\

This leads to the possibility to extract an infinite amount of energy from the movement of the balance. In fact, introducing a friction needed to extract energy, one initially observes a decrease in the amplitude of the oscillations, but there would be no continuous damping of the oscillations of the structure: after each cycle, the photon looses no energy. This implies that \textit{the first principle of thermodynamics is violated}. A similar violation occurs in the thought experiment conceived by H. Bondi \cite{bo}: in this case the inconsistency with the principle of energy conservation was resolved by talking into account the redshift of the photon when travelling upwards in a gravitational field. In our thought experiment, the redshift that the photon would undergo when travelling back, say from D to C in fig. 2, would cause only a finite reduction of the photon energy. After each cycle, when the photon returns to the lowest part D of the balance, it regains the reshifted energy. Additionally, one can restrict the gravitational redshift by limiting the rotation angle around the fulcrum of the balance suitably. 
\

\subsection{Photon redshift}
For a spherical body of mass M, such as our fictious planet, we find the following relation between the shift in frequency of the photon and the gravitational potential difference during the travel \cite{ch}:

\begin{equation*}
 \frac{\nu_{r}-\nu_{e}}{\nu_{e}} = - \frac{G_N M}{c^2} \left( \frac{1}{r_r} - \frac{1}{r_e} \right)
\end{equation*}
\

Suppose that the altitude and the width of the balance have the same value as the planet radius $R$.  A quick calculation, considering a planet with the mass of the Sun ($M \simeq 1.98874 \cdot 10^{30}\ kg$) but with the radius of the Earth ($R \simeq 6371\ km$) and non-rotating, leads to

\begin{equation*}
 \frac{\nu_{r}-\nu_{e}}{\nu_{e}} = \frac{G_N\ 1.98874 \cdot 10^{30}}{c^2} \left(\frac{1}{14246}-\frac{1}{10073}\right)
\end{equation*}

so that

\begin{equation*}
 1- \frac{\nu_r}{\nu_e} = 0.0429476
\end{equation*}
\

In an ideal system as the one described above, such a non-zero ratio between frequencies signals a \textit{periodic variation of the gravitational mass distribution in the scale}, leading to an oscillation of the same.
\
\\
 
Therefore, it appears that gravitational redshift is not sufficient to prevent violation of the first principle of thermodynamics. 
Furthermore, \textit{a violation of the second principle of thermodynamics} occurs when the scale is placed in a linearly accelerated frame. Imagine that some sufficiently hot body looses heat by emitting light. A given scale could absorb a single photon. The possibility to observe complete oscillations of the balance shows that all the absorbed heat has as unique effect the production of work. Thus one has a contradiction with Kelvin's formulation of the second law of thermodynamics.   

\section{Tidal effects and rotating system}

\subsection{Tidal effect}
Let us now examine what would happen if the scale were placed in a gravitational field characterised by \textit{non-vanishing tidal components}.
If we assume that the scale is at a given instant of time in the configuration depicted in fig. 2, and that the gravity is sufficiently weak to allow the use of a Newtonian approximation in a an almost Euclidean space. 
Tidal effects can prevent oscillations of the balance if the gravitational acceleration at the lower level D is greater than (or equal to) the acceleration at the higher level C, even if the latter part of the balance contains the travelling photon and possesses a greater total gravitational mass. 
If $m_D$ and $m_C$ are respectively the gravitational masses of part D and C of the scale and if $g_C$ and $g_D$ are respectively the gravitational accelerations at the levels of C and D, then the oscillation will not occur if 

\begin{equation}
 m_D g_D  >  m_C g_C
\end{equation}
\

Since the coil C contains the photon, its mass can be expressed as: 

\begin{equation*}
 m_C = m_D + \frac{h \nu}{c^2}
\end{equation*}
\

with $\nu$ being the photon frequency and c the speed of light.  
\

Rewriting the acceleration $g_D$ as $g_D = g_C + \Delta g$, with $\Delta g = g_D - g_C$, the inequality becomes 

\begin{equation*}
 m_D ( g_C + \Delta g  )  > \left( m_D + \frac{h \nu}{c^2} \right) g_C 
\end{equation*}
\

which leads to 

\begin{equation}
 m_D \Delta g > \left( \frac{h \nu}{c^2} \right) g_C
\end{equation}
\

The frequency can be expressed as as $\nu = 1/ T$, with $T$ being the period, and the acceleration $g_C$ can be written as 

\begin{equation*}
 g_C = \frac{GM}{R^2}
\end{equation*}
\

where $M$ is the mass of the body that generates the gravitational field and $R$ is the distance from the coil C to the center of the body.
\
\\

As a result the inequality (2) becomes 

\begin{equation*}
 \Delta g\ T > \frac{h G}{c^2} \frac{1}{R^2} \frac{M}{m_D}
\end{equation*}
\

The Planck area is defined as $A_{pl} = \frac{Gh}{2 \pi c^3}$, and $R^2$ can be expressed in terms of the area A of the surface of a  sphere of radius $R$ as $R^2 = A/4\pi$.  With these notations the above inequality becomes 

\begin{equation}
 \Delta g\ T > 8\pi^2 \frac{A_{pl}}{A} \frac{M}{m_D} c
\end{equation}
\

Therefore, according to (3), given the photon period, the distance from the lowest part of the balance to the centre of a gravitation-generating body and given the masses of the coils of the scale and of the source of the gravitational field, \textit{the relative gravitational acceleration between the levels of two coils of the balance can not be arbitrarily small}, otherwise a violation of the principles of thermodynamics might occur. 
Furthermore, \textit{from the occurrence of the constants $h$, $G$ and $c$ in the inequality (3), it appears that the origin of this constraint on the value of tidal effects might be linked to quantum effects in a gravitational field}. More precisely, equation (3) is reflecting a quantum gravity effect, in the same sense that Hawking's black hole entropy formula \cite{ha} reflects properties of quantum gravity, because of the simultaneous occurrence of the constants h, G and c. Indeed, if the factor $h G/c^3$ tends to zero, then the minimal tidal effect requirement would not be applicable and the basic result of the paper would break down in this limit, which includes:

\begin{itemize}
 \item special relativity (G and h tend to zero)
\\
 \item classical general relativity (h tends to zero)
\\
 \item classical Newtonian gravity (h tends to zero and c tends to infinity)  
\\
 \item relativistic quantum physics  (only G tends to zero, while h remains non-zero and c finite)
\\
 \item non-relativistic quantum mechanics in a Newtonian gravitational field (c tends to infinity).
\end{itemize}

\

\subsection{Rotating system}
Linearly accelerated frames can not produce tidal effects, required in order to prevent violations of the principles of thermodynamics. 
\

Imagine that the scale is placed on a rotating disk or platform, as in fig.(4).  When the balance starts oscillating such that coil D is at greater distance from the centre of rotation than coil C, there exists a greater centrifugal acceleration at D when compared to its value at C.
\
\\

Therefore,  the equivalent gravitational field experienced by an observer in the rotating frame has non-vanishing tidal effects, and these can prevent oscillations of the balance: however, for any given angular velocity of the frame, \textit{the relative acceleration between the two coils of the balance can not be arbitrarily small}.  This implies that the distance between the two extreme oscillation levels of the balance can not be arbitrarily small: for a given rotating disk, this means that the distance between a point and the centre of the disk can vary only discontinuously. Thus, according this heuristic picture, \textit{the radial coordinate of the rotating disk is quantised}.
\
\\
  
\begin{center}
\begin{figure}
 \includegraphics[width=180pt]{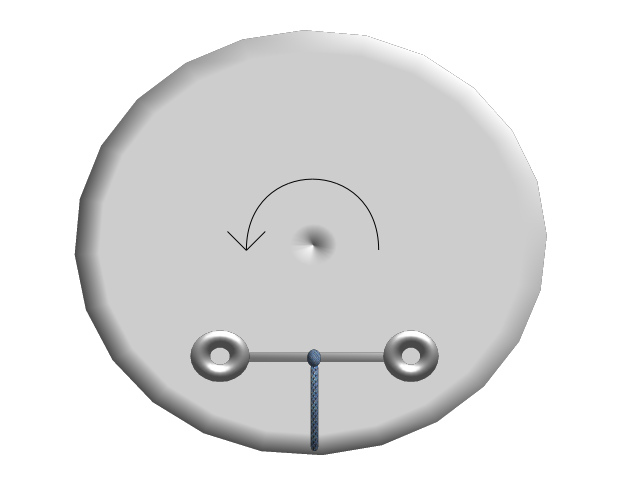}
 \caption{The scale is placed in a rotating disk.}
\end{figure}
\end{center}

\section{Conclusions}

The thought experiment considered in this work generalises an ideal experiment discussed by H. Bondi \cite{bo}.
\

It appears, however, as far as our ideal experiment is concerned, that gravitational redshift is not sufficient to prevent the violations of the first and the second principles of thermodynamics.  In particular, these contradictions can occur if a gravitational field is considered with no tidal effects, or if the experiment is performed in a linearly accelerated frame. On the other hand, if minimal tidal effects  inherent to a gravitational field are taken into account, then the contradictions with the principles of thermodynamics can be avoided. Tidal effects can be experienced on a rotating disk, and in such frame no violations would occur if the radial separation between two points can not be arbitrarily small, since the relative acceleration depends on the distance from the centre of rotation. This shows that there is a physically profound difference between linearly accelerated frames and rotating frames and that only in the latter the inertial effects mimic consistently a gravitational field. This physical distinction between rotation and other motions has also been highlighted in the work of Gr\o{}n \cite{gr}, in which the fundamental question on the relativity of rotation has been addressed.  
\

The heuristic analysis of the tidal effect in a gravitational field shows through inequality (3) that the necessary existence of a minimal non-vanishing relative gravitational acceleration might be related to quantum gravitational effects. In any case, since the relative acceleration depends on the spatial separation between  points of space, one would expect that these distances can not be arbitrarily small in order to guarantee non-vanishing tidal forces.    
\
\\

The heuristic arguments developed in this paper might also be related to the Weyl Curvature Hypothesis (WCH) of R. Penrose \cite{pe}. The Weyl curvature essentially describes the tidal effects of a gravitational field, and if some invariant function constructed with the components of the Weyl tensor is linked to the gravitational entropy, then the WCH states that the Weyl curvature associated to a cosmological space-time should have a very small value initially, close to the Big Bang singularity, whereas at late-times in the far future the solution ought to be characterised by a large value of the Weyl curvature. The cosmological space-time solution exhibits in this way a time-asymmetric evolution, compatible with the second principle of thermodynamics.
\
\\

Our analysis suggests that tidal effects are necessary and essential features of gravity, which guarantee compatibility with the principles of thermodynamics. Thus the evolution of cosmological space-time solutions possessing such features will exhibit a time-asymmetric evolution implied by the second principle of thermodynamics.  As R. Penrose argues \cite{pe}, this time-asymmetry might ultimately be related to quantum gravity.

\end{document}